\documentclass[12pt,preprint]{aastex}

\usepackage{amsmath}

\usepackage{graphicx}
\usepackage{multirow}
\usepackage{color}
\usepackage{natbib}

\begin{document} 

\title{The effect of orbital evolution on the Haumea (2003 EL$_{61}$) collisional family}

\author{Kathryn Volk and Renu Malhotra}
\affil{Lunar and Planetary Laboratory, The University of Arizona, Tucson, AZ 85721, USA.}
\email{kvolk@lpl.arizona.edu}

\begin{abstract}

The Haumea family is currently the only identified collisional family in the Kuiper belt.  We numerically simulate the long-term dynamical evolution of the family to estimate a lower limit of the family's age and to assess how the population of the family and its dynamical clustering are preserved over Gyr timescales. We find that the family is not younger than 100 Myr, and its age is at least 1 Gyr with 95\% confidence.  We find that for initial velocity dispersions of $50-400$ ms$^{-1}$, approximately $20-45$\% of the family members are lost to close encounters with Neptune after 3.5 Gyr of orbital evolution.  We apply these loss rates to two proposed models for the formation of the Haumea family, a graze-and-merge type collision between two similarly sized, differentiated KBOs or the collisional disruption of a satellite orbiting Haumea.  For the graze-and-merge collision model, we calculate that $>85\%$ of the expected mass in surviving family members within $150$~ms$^{-1}$ of the collision has been identified, but that one to two times the mass of the known family members remains to be identified at larger velocities.  For the satellite-break-up model, we estimate that the currently identified family members account for $\sim50\%$ of the expected mass of the family.  Taking observational incompleteness into account, the observed number of Haumea family members is consistent with either formation scenario at the $1 \sigma$ level, however both models predict more objects at larger relative velocities ($>150$~ms$^{-1}$) than have been identified.

\end{abstract}

\section{Introduction}\label{s:Introduction}

The Haumea (2003 EL$_{61}$) collisional family was discovered by \citet{Brown2007} who noted that Haumea and five other Kuiper belt objects (KBOs) shared a spectral feature that is indicative of nearly pure water ice on the surfaces of the bodies.  These six KBOs, along with four additional family members identified by \citet{Schaller2008}, \citet{Snodgrass2010}, and \citet{Ragozzine2007}, can all be dynamically linked to Haumea, and there do not appear to be any dynamically unrelated KBOs that share this spectral feature.  Aside from being spectrally linked to these other KBOs, Haumea itself shows signs of its collisional past.  Despite having a nearly pure water ice surface, Haumea's density is $\sim2.6$ g cm$^{-3}$ \citep{Rabinowitz2006}, which is higher than expected for typical assumed ice/rock ratios in the Kuiper belt \citep{Brown2008}; one way to achieve this higher density is to have a catastrophic collision between a differentiated proto-Haumea and another KBO in which proto-Haumea loses a substantial fraction of its water ice mantle \citep{Brown2007}.  This scenario is supported by the presence of at least two water ice satellites \citep{Barkume2006,Ragozzine2009}.  Haumea also has an elongated shape and a very short spin period of $\sim4$ hours that is unlikely to be primordial \citep{Rabinowitz2006,Lacerda2007}.

\citet{Ragozzine2007} examined the dynamical connections between the identified Haumea family members.  These connections are made by first estimating the orbit of the center of mass of the colliding bodies, and then estimating the ejection velocities of each family member relative to the collision's center of mass.  The ejection velocity is given by
\begin{equation}\label{eq:dv}
\Delta \vec{v}= \vec{v} - \vec{v}_{cm}
\end{equation}
where $\vec{v}_{cm}$ is the estimated collision's center-of-mass velocity.  Because Haumea is by far the largest remnant from the collision, its orbit immediately after the collision should have nearly coincided with the center-of-mass orbit. However, Haumea is currently located at the boundary of the 12:7 mean motion resonance (MMR) with Neptune; over long timescales, the chaotic zone of this resonance causes a random walk of the proper elements such that Haumea's current orbit may be significantly distant from its post-collision orbit.  \citet{Ragozzine2007} estimate the center-of-mass collision orbit by minimizing the sum of the relative speeds of all family members, assuming that Haumea's semimajor axis and its Tisserand parameter with respect to Neptune are both conserved during its chaotic evolution; they then use Haumea's present distance from the collision's center-of-mass orbit, together with a calculation of its chaotic diffusion rate, to estimate the age of the collisional family to be $3.5\pm2$ Gy.  Given the exceedingly low collision probabilities for objects large enough to form the Haumea family in the current Kuiper belt, the family is likely to be old.  However, the family probably cannot have formed in the primordial, much more massive Kuiper belt, because whatever caused the mass of the Kuiper belt to be depleted (by an estimated 2 or 3 orders of magnitude) would have also destroyed the dynamical coherence of the family~\citep{Levison2008}.  The high inclination ($\sim27^{\circ}$) of the family also argues against a primordial origin, because such large inclinations are probably products of the excitation and mass depletion of the Kuiper belt.  Thus, it appears that the Haumea family-forming collision occurred near the end of the primordial, high-mass phase of the Kuiper belt.

Several of the largest KBOs show evidence of their collisional past (see review by \citet{Brown2008}), but the Haumea family is the only collisional family that has been identified in the Kuiper belt.  The dynamical connections between the members of the family allow us to place some constraints on the type of collision that formed the family and also constrain the age of the family as being old, but probably not primordial.  These characteristics  make the Haumea family an excellent probe of the collisional environment in the Kuiper belt following the excitation and mass depletion event; understanding the type of collision that created the family (especially the relative sizes and speeds of the impactor and target) would provide valuable insight into the size and orbital distribution of the Kuiper belt at the time of the collision (see discussions of this in \citet{Marcus2011} and \citet{Levison2008}).

Proposed models for the formation of the Haumea family have attempted to reproduce the family's relatively small velocity dispersion ($\sim150$~ms$^{-1}$) and to explain the compositional and orbital characteristics of the family.  However, the orbits of the family members have been sculpted by several gigayears of dynamical evolution.  In this paper we use numerical simulations to determine how this orbital evolution affects the dynamical coherence of the family. In Section~\ref{s:sims}, we determine the loss rates for the family, which depend on the initial velocity dispersion from the collision, and we determine how the velocity dispersion of the surviving family members is altered over time; from these simulations, we also obtain a hard lower limit for the age of the family.  In Section~\ref{s:formation_models}, we apply these results to the family-formation models of \citet{Leinhardt2010} (a graze-and-merge type collision between two similarly sized, differentiated KBOs) and \citet{Schlichting2009} (the collisional disruption of a satellite orbiting Haumea), and we compare the predictions from these two formation models to the current observations of the family.  Section~\ref{s:conclusions} provides a summary of our results and conclusions.

\section{Orbital evolution of the Haumea family}\label{s:sims}

Even though the identified Haumea family members (see Table~\ref{t:known_family}) have a fairly low velocity dispersion ($\Delta v \sim 150$~ms$^{-1}$), their proper orbital elements span a relatively large range in semimajor axis, $a$, and eccentricity, $e$, (a range that is typical of classical KBOs), and they have atypically large inclinations, $i$, of $\sim27^{\circ}$.  Using the data for their best-fit orbits\footnote{orbit information was taken from the AstDyS website (http://hamilton.dm.unipi.it/astdys)}, we did a 10 Myr numerical simulation to obtain the average values of $a,e$ and $i$ for each family member over that time span, and we calculated the corresponding values of $\Delta v$ (equation~\ref{eq:dv}) relative to the center-of-mass collision orbit determined by~\citet{Ragozzine2007}; these are listed in Table~\ref{t:known_family} for the family members identified by \citet{Brown2007}, \citet{Schaller2008}, \citet{Ragozzine2007}, and \citet{Snodgrass2010}.  Below, we examine the orbital distribution of the known family members to refine the center-of-mass orbit in light of the additional identified family members since \citet{Ragozzine2007}.   We use the results of long-term numerical simulations to estimate how much the family's orbits have evolved since its formation, and we obtain a hard lower limit on the age of the family. 

\subsection{Collision center-of-mass orbit and a lower limit on the family's age}\label{ss:cmorbit}

\begin{figure} 
   \centering
   \includegraphics{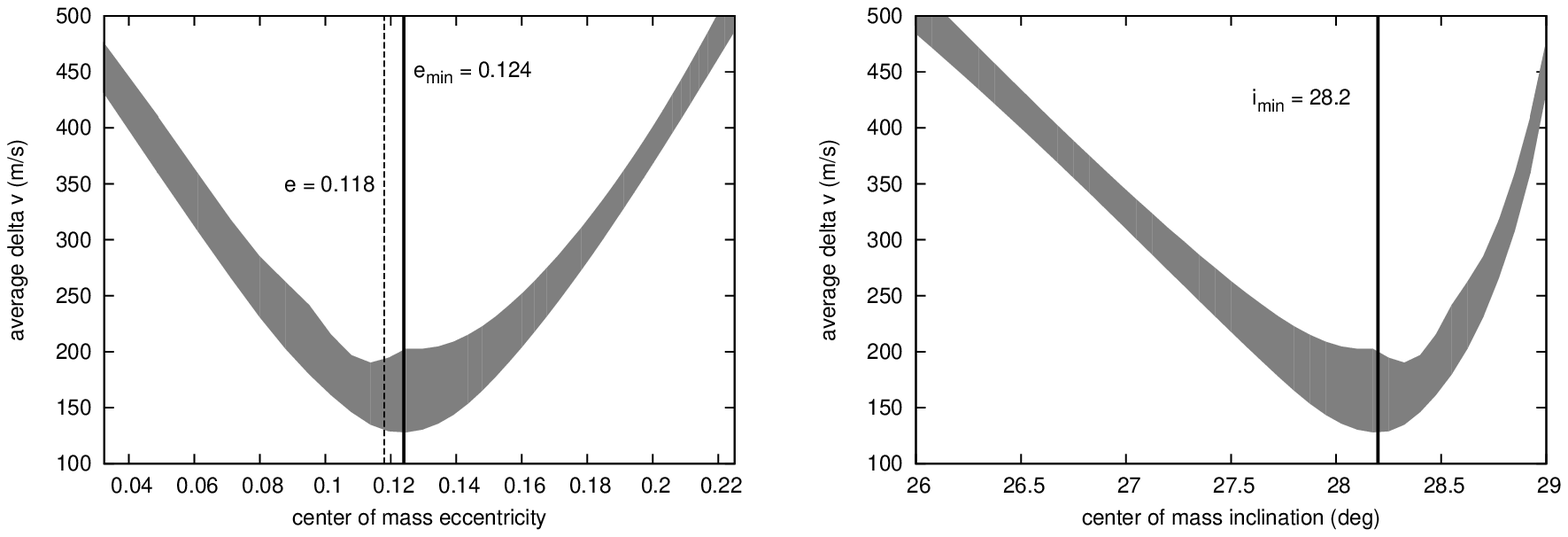}
   \caption{Average $\Delta v$ for the nine identified Haumea family members as a function of the eccentricity and inclination of the collision center-of-mass orbit.  The shaded regions show the range of $\Delta v$ over all values of the argument of pericenter ($\omega$) for the center-of-mass orbit (assuming the value of the mean anomaly that minimizes $\Delta v$ for each value of $\omega$).}
   \label{f:cm_orbit}
\end{figure}

We use the average values of $a, e,$ and $i$ for the nine identified family members (Table~\ref{t:known_family}) to re-calculate the center-of-mass collision orbit using the method described by \citet{Ragozzine2007}:  we minimize the sum of $\Delta v$ for the nine family members while fixing the semimajor axis of the center-of-mass orbit at that of Haumea's current orbit, allowing its eccentricity and inclination to vary such that Haumea's current Tisserand parameter with respect to Neptune ($T_N = 2.83$) is maintained, and allowing the mean anomaly, $M$, and the argument of pericenter, $\omega$, to vary freely;  the longitude of ascending node, $\Omega$, is ignorable as it does not affect the distribution of $\Delta v$.  Figure~\ref{f:cm_orbit} shows the results of this calculation for a range of eccentricity and inclination combinations of the collision center-of-mass orbit.  The lower limit of the shaded region in the figure is the value of the family's average $\Delta v$ found by selecting values of the mean anomaly and argument of pericenter that minimize $\Delta v$; the shaded area shows the range in $\Delta v$ obtained by allowing $\omega$ to vary, but still selecting the value of $M$ that minimizes $\Delta v$ for each value of $\omega$.  Parameters along the lower boundary of the shaded regions represent collisions occurring very near to the ecliptic plane, while parameters along the upper boundary represent collisions at the extreme, off-ecliptic points in the orbit ($\sim15-20$ AU above the ecliptic plane).  The difference in average $\Delta v$ for the different values of $\omega$ is a factor of $\sim 2$, as noted by \citet{Ragozzine2007}; this increase in average $\Delta v$ for off-ecliptic collision points is due to the fact that producing the observed family's spread in inclination requires a larger $\Delta v$ at these locations.  Because collisions near the ecliptic are much more probable than off-ecliptic collisions, we choose the center-of-mass orbit that minimizes the lower portion of the filled curve in Figure~\ref{f:cm_orbit}.  The result is $(a,e,i,\omega,M) = (43.1$ AU$, 0.124, 28.2^{\circ}, 270^{\circ}, 76^{\circ})$.  This is very similar to the collision center-of-mass orbit determined by \citet{Ragozzine2007}:  $(a,e,i,\omega,M) = (43.1$ AU$, 0.118, 28.2^{\circ}, 270.8^{\circ}, 75.7^{\circ})$, indicating that the newer family members do not significantly affect the estimate of the collision center-of-mass orbit. The small difference in the eccentricity does not much affect the values of $\Delta v$ for the family members (both values of $\Delta v$ are listed in Table~\ref{t:known_family}) because the calculated $\Delta v$ is a fairly flat function of eccentricity within $\pm\sim10\%$ of its minimum.  

In the above calculations, as in~\citet{Ragozzine2007}, we assumed a constant semimajor axis and conservation of the Tisserand parameter during the chaotic evolution of Haumea's orbit.  To test the validity of this assumption, we performed numerical simulations of resonant diffusion within the 12:7 MMR, and we find that Haumea's Tisserand parameter can vary by $\pm0.5\%$.  This is a small variation, but it does affect the allowable combinations of $e$ and $i$ for the best-fit center-of-mass orbit.  We performed the minimization of the sum of $\Delta v$ for the identified family members while allowing $e$ and $i$ to vary independently, and we find a slightly revised best-fit center-of-mass orbit:  $(a,e,i,\omega,M) = (43.1$ AU$, 0.124, 27.3^{\circ}, 276^{\circ}, 70^{\circ})$.  This orbit has a Tisserand parameter $T_N = 2.84$, which is within the range of $T_N$ found in our numerical simulations.  If we additionally relax the constraints to allow the semimajor axis of the orbit to vary by $\pm0.15$ AU (the approximate range of variation in the 12:7 MMR), we find very similar results:  $(a,e,i,\omega,M) = (43.1\pm0.15$ AU$, 0.121, 27.3^{\circ}, 278^{\circ}, 68^{\circ})$.  These alternate minimum $\Delta v$ center-of-mass orbit fits give us an estimate of the uncertainties in the orbital parameters:
\begin{equation*}
(a,e,i,\omega,M)_{cm} = (43.1 AU, 0.115-0.132, 27-28.3^{\circ}, 270-278^{\circ}, 68-76^{\circ}).
\end{equation*}    
These orbits all represent collisions near the ecliptic plane, and the \citet{Ragozzine2007} estimate falls within the uncertainties.

We can use the collision center-of-mass orbit to set a lower limit on the age of the Haumea family by determining the minimum time necessary for such an orbit to diffuse to Haumea's current eccentricity ($e=0.19$) in the 12:7 MMR with Neptune.  We generated 800 test particles with initial conditions within the uncertainties of the collision center-of-mass orbit found above, randomized their initial mean anomaly, and integrated these for 1 Gyr.  We find from this simulation that $\sim3\%$ and $\sim6\%$ of the test particles reach Haumea's eccentricity by 500 Myr and 1 Gyr, respectively; this is a slightly lower efficiency than \citet{Ragozzine2007} found from similar simulations ($10\%$ had diffused by 1 Gyr), but the two results are consistent given that the number of test particles in their simulation was only 78.  These results allow us to conclude with $\sim95\%$ confidence that the Haumea family is older than 1 Gyr.  The fastest that any test particle in our simulation diffused to Haumea's eccentricity was $\sim100$ Myr (Fig.~\ref{f:haumea});  this indicates that 100 Myr is a strong lower limit on the age of the family.

Another way to estimate the lower limit on the family's age is to examine the precession of the orbital planes of the identified family members.  The current values of the longitudes of ascending node, $\Omega$, of the known family members are indistinguishable from a random distribution.  Imediately after the family forming collision, the family members will share a common line of nodes on the collision center-of-mass orbit plane, but will have different values of the other orbital elements.  After the collision, the differences in semimajor axes amongst the family members cause their orbit planes to precess at slightly different rates.   The nodal precession rates range from $64^{\circ}Myr^{-1}$ to $81^{\circ}Myr^{-1}$ for the family members strongly affected by MMRs with Neptune and $69^{\circ}Myr^{-1}$ to $72^{\circ}Myr^{-1}$ for the non-resonant family members (as determined from our numerical simulations of their best-fit orbits).  Considering the differences in these rates, we expect the nodes to be randomized on a 20 Myr timescale for the resonant family members and a 100 Myr timescale for the non-resonant family members.  Both this estimate and the resonant diffusion timescale of Haumea's eccentricity within the 12:7 MMR indicate that 100 Myr is a hard lower limit on the age of the family.

\subsection{Long-term orbital evolution}
 
The Haumea family members' range in semimajor axis includes regions affected by various MMRs with Neptune.  This means that since the time of the collision (at least 100 Myr ago, but most likely several Gyr ago), the orbital distribution of the family has been modified by dynamical evolution, and some family members have probably been removed by orbital instabilities.  Comparisons of formation models to the current set of observed family members must account for this orbital modification and decay of the total population.  

To determine the effects of long-term orbital evolution on the family, we performed a suite of eight numerical simulations, each with a cloud of 800 test particles representing family members generated with an isotropic distribution of initial ejection velocity vectors about the collision center-of-mass determined by \citet{Ragozzine2007}:  $(a,e,i,\omega,M) = (42.1$ AU$, 0.118, 28.2^{\circ}, 270.8^{\circ}, 75.7^{\circ})$.  (As discussed in Section~\ref{ss:cmorbit}, there is some uncertainty in this collision center-of-mass orbit, but the simulation results do not strongly depend on the exact values chosen, as all of the allowed range results in test particles spread over very similar ranges in $a$, $e$, and $i$.)   In each of the eight simulations, we adopted different values of the magnitude, $\Delta v$, of the initial ejection velocity of the cloud of test particles: $\Delta v = 50,100,150,..., 400$ ms$^{-1}$, respectively.  We then integrated these test particle clouds forward in time for $4$ Gyr under the influence of the Sun and the four outer planets (in their current configuration), using the symplectic orbital integration method of \citet{Wisdom1991}.  Any test particles that approached within a Hill sphere of Neptune were considered lost, because these are unstable on very short timescales.

\begin{figure} 
   \centering
   \includegraphics{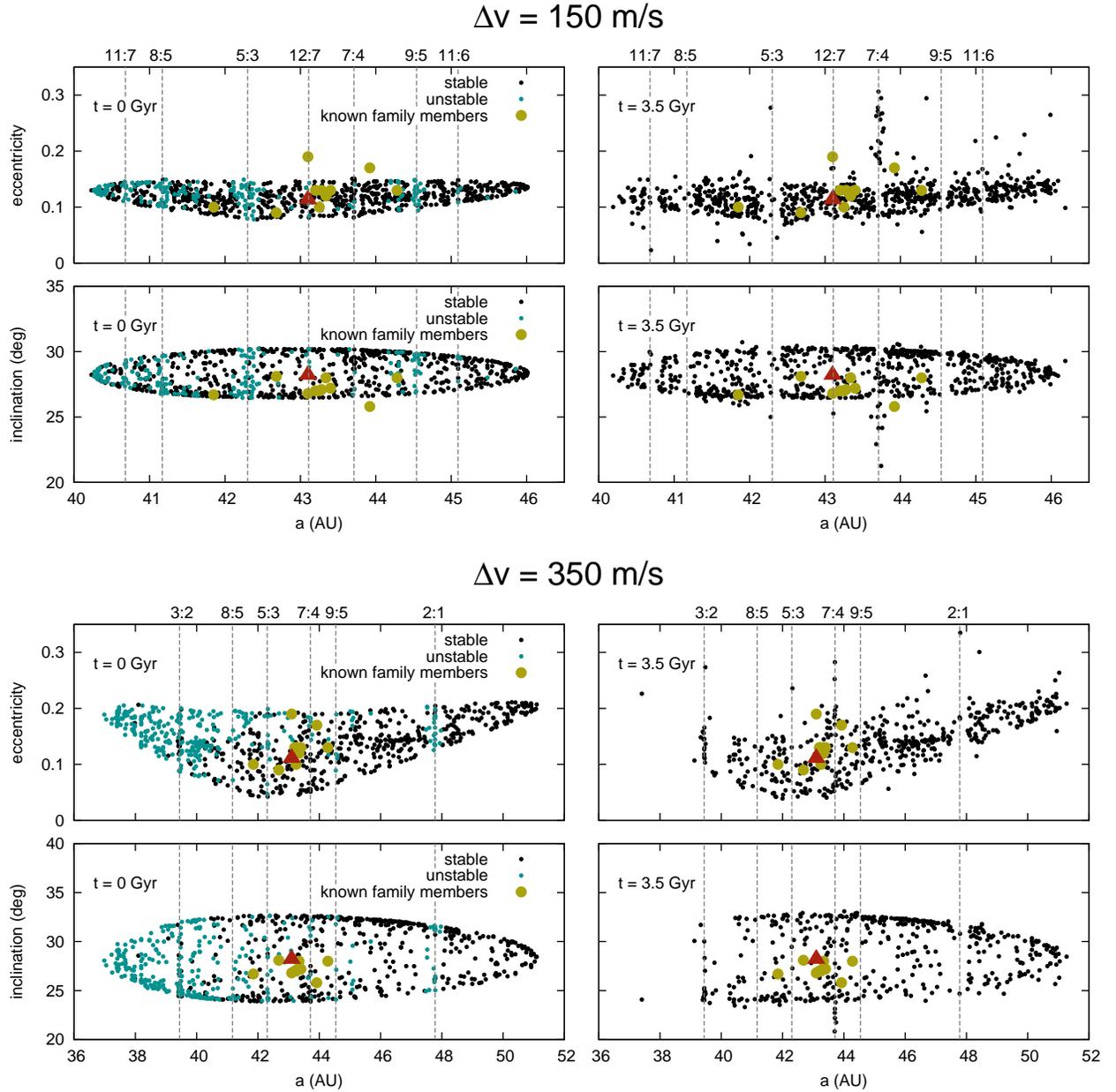}
   \caption{{\footnotesize Proper eccentricity and semimajor axes for the observed family members (green circles) and for sets of test particles with an isotropic $\Delta v = 150$~ms$^{-1}$ (top four panels) and $\Delta v = 350$~ms$^{-1}$ (bottom four panels) from the collision center-of-mass orbit (red triangle).  Black points indicate test particles that are stable over 3.5 Gyr; blue points indicate test particles that are unstable and are removed via close encounters with Neptune within 3.5 Gyr.  The locations of various MMRs with Neptune are indicated.  Note that the scaling of the x- and y-axes is different in the top four and bottom four panels.  The two values of $\Delta v$ are shown as examples of the evolution of the family.  The simulations for $\Delta v = 50,100,200,250,300$ and $400$~ms$^{-1}$ are not shown.}}
   \label{f:snapshots}
\end{figure}

\begin{figure} 
   \centering
   \includegraphics{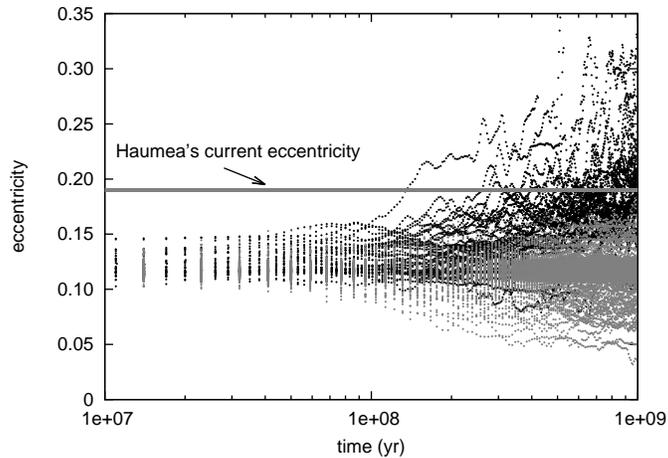}
   \caption{Eccentricity time evolution for 800 clones of the best-fit collision center-of-mass orbit.  The grey dots are for particles which have not evolved to Haumea's current eccentricity by the end of the 1 Gyr simulation (94\% of the cloned orbits).  The black dots are the orbits which do reach Haumea's eccentricity (6\% of the cloned orbits).  The minimum time required to diffuse to Haumea's eccentricity is just over 100 Myr.}
   \label{f:haumea}
\end{figure}

Figure~\ref{f:snapshots} plots  the $a-e$ and $a-i$ distributions for two of our simulations; both the initial distributions and a snapshot at 3.5 Gyr are shown.  In these plots, the test particles are color coded according to their stability (a particle is considered to be unstable if it approaches within a Hill sphere of Neptune at any point in the simulation); most of the test particles that are unstable are located either near the inner edge of the family (where their initial perihelion distances are nearly Neptune crossing) or near the labeled MMRs with Neptune.  Previous studies have found that this region of the Kuiper belt is strongly affected by even fairly high order MMRs \citep{Chiang2003,Lykawka2007,Volk2011}, so it is not surprising that the family is strongly affected as well.  The effect of the resonances on the unstable test particles is to increase their eccentricity until they become Neptune crossing and then have close encounters with Neptune.  A few percent of the test particles survive in resonance until the end of the simulation; this result is consistent with the resonant fraction found by \citet{Lykawka2012} in a similar study of the evolution of the Haumea family.  These stable resonant test particles are mostly found in the 3:2 and 7:4 MMRs; in these cases, the long-lived test particles are additionally stabilized due to the Kozai resonance \citep{Kozai1962}.  The Kozai resonance causes a test particle's argument of perihelion to librate about $90^{\circ}$ which ensures that perihelion occurs well away from the ecliptic plane, protecting the test particle from close encounters with Neptune.  

\begin{figure} 
   \centering
   \includegraphics{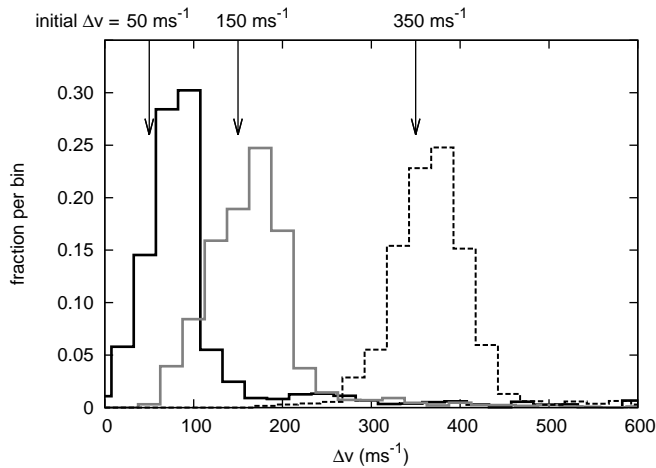}
   \caption{The distribution of relative velocity, $\Delta v$, for the surviving test particles at 3.5 Gyr in the two numerical simulations shown in Figure~\ref{f:snapshots} as well as the simulation with initial $\Delta v = 50$ ms$^{-1}$;  in each simulation we started with a cloud of particles having the same magnitude of $\Delta v$ (of 50, 150, and 350~ms$^{-1}$), but isotropically distributed in direction, relative to the velocity of the collision center-of-mass orbit.}
   \label{f:deltav}
\end{figure}

The fraction of test particles that survive as family members up to $1.5$ and $3.5$ Gyr are listed in Table~\ref{t:survival_rates} for each value of $\Delta v$; for initial $\Delta v$ of $50-200$~ms$^{-1}$ (typical of the observed family) $20-25\%$ of the family is lost by $3.5$ Gyr, which is consistent with the \citet{Lykawka2012} results.  In addition to the erosion of the population, we also find that the dynamical clustering in proper elements grows weaker (and the apparent $\Delta v$ of the test particles increases) over the course of the simulations.  We determined the $\Delta v$ for each test particle by calculating its proper $a, e$ and $i$ (taking the average over the last 50 Myr of the simulation) and then allowing the values of the orbital angles to vary until we find the smallest difference between the test particle's orbital velocity and the orbital velocity of the collision's center-of-mass orbit.  We find that the chaotic diffusion in orbital elements induces a spread of $50-100$~ms$^{-1}$ among the test particles and shifts the average to slightly higher than their initial value of $\Delta v$.  Figure~\ref{f:deltav} shows the distributions of $\Delta v$ at 3.5 Gyr for three simulations in which the initial $\Delta v$ had values of $50$~ms$^{-1}$, $150$~ms$^{-1}$, and $350$~ms$^{-1}$.  We conclude that, while some individual test particles that are strongly affected by MMRs with Neptune experience larger changes in $\Delta v$, for the non-resonant Haumea family members, it is likely that the value of $\Delta v$ they acquired at the time of the collision was within $\pm50-100$~ms$^{-1}$ of their current $\Delta v$.

\section{Implications for family formation}\label{s:formation_models}

The numerical simulations described above show how long-term orbital evolution will affect the Haumea family's dynamical clustering.  In this section, we examine the implications of those results for the proposed family formation models of \citet{Leinhardt2010} and \citet{Schlichting2009}.  These models make specific predictions for the mass and velocity distribution of the family members immediately following the formation of the family.  We use our simulations of the family's orbital evolution to evolve the models' predicted mass and velocity distributions to the current epoch so we can compare the models' predictions to the currently observed family.

\subsection{Overview of proposed formation models}\label{ss:proposed_models}

\citet{Brown2007} estimated that the Haumea collisional family was created in a catastrophic collision between a proto-Haumea of radius $R\sim830$ km and another KBO with a radius of $\sim500$ km.  However, the low velocity dispersion amongst the observed family members is problematic for such a model. In a catastrophic collision between two large KBOs, the velocity dispersion of the family members should be close to the escape velocity of the largest remnant, $\Delta v \sim v_{\rm{esc,Haumea}} \sim 900$~ms$^{-1}$ (see discussion in \citet{Leinhardt2010} and \citet{Schlichting2009}); in contrast, the observed family has $\Delta v = 50$--$300$~ms$^{-1}$. 

To explain the low velocity dispersion of the observed family, \citet{Schlichting2009} propose that the family originates from the catastrophic disruption of a satellite orbiting Haumea, rather than the disruption of a proto-Haumea.  This actually requires two different collisions: a collision between a proto-Haumea and another large KBO that creates a large, icy satellite and gives Haumea its unusual shape and fast spin, then a subsequent collision between the satellite and another KBO.  The latter would create a collisional family with values of $\Delta v$ close to the escape speed of the satellite rather than of Haumea.  Assuming a primarily water ice composition, these authors estimate that the disrupted satellite would have had a radius $R\sim 260$ km to account for the mass of Haumea's remaining satellites and the rest of the collisional family; the expected $\Delta v$ would be $\sim 200$~ms$^{-1}$ for a satellite of this size.  For a collisional family formed in this way, the authors estimate that the total mass of the family at formation would be no more than $\sim5\%$ of the mass of Haumea ($M_H \approx 4.2 \times 10^{21}$ kg).  

\citet{Leinhardt2010} propose an alternative formation mechanism for the family:  a graze and merge collision event between two similarly sized, radius $R\sim850$ km, differentiated KBOs.  In this scenario, the two impacting bodies merge after the collision, resulting in a very fast rotating object.  The family members and satellites are a result of mass shedding due to the high spin rate of the merged body (Haumea) rather than being direct impact ejecta; this accounts for the low velocity dispersion of the family members. The authors ran several collision simulations and found that the resulting family would have a mass of $\sim4-7\%$ $M_H$, most of which would be found within $\sim300$~ms$^{-1}$ of the collision orbit.

\subsection{Dynamical evolution of the family}\label{ss:dynamical_evolution}

To determine how the dynamical evolution of the family will affect the mass and $\Delta v$ estimates from these formation scenarios, we generate synthetic families for the two models.  We detail the assumptions for each step below, but the overall procedure is as follows:  we first sample the size distribution predicted for the collisional model to generate a list of family members and their sizes.  We then assign these members an initial $\Delta v$ according to the distribution of velocities from the model.  This creates a snapshot of the family immediately after formation.  To account for 3.5 Gyr of orbital evolution, we randomly assign an orbital history from one of our eight simulations (Section~\ref{s:sims}) to each of our synthetic family members (assigning it from the appropriate simulation based on initial $\Delta v$).  Having assigned each family member an orbital history, we then take a snapshot of the surviving family members' orbital element and mass distributions at 3.5 Gyr.  From this we can calculate the expected mass vs. $\Delta v$ distribution of the family for each model.  

For the graze-and-merge collisional model, \citet{Leinhardt2010} provide plots of the cumulative number of collisional fragments as a function of fragment mass and the cumulative mass of fragments as a function of $\Delta v$ from each of their high-resolution collision simulations (their Figure 3).  We use the given total mass of collisional fragments to convert the normalized number of fragments presented in their Figure 3a to an absolute number of fragments.  To convert the cumulative mass distribution to a cumulative size distribution, we assume that all the fragments have a uniform density of 1.15 g cm$^{-3}$, which is consistent with the family being composed of 80\% water ice by mass (their Table 2).  We construct a set of synthetic family members from this size distribution and assign each fragment an initial $\Delta v$ based on the mass vs. velocity distribution (their Figure 3b); each bin in $\Delta v$ is filled with randomly selected family members until the specified mass in that bin is reached.  Based on its initial value of $\Delta v$, each family member is then randomly assigned a 3.5 Gyr orbital history from our numerical simulations; in this way some family members are removed from the population, and the others diffuse in orbital elements and apparent $\Delta v$.   \citet{Leinhardt2010} detail four different successful graze-and-merge collision simulations with slightly varying initial conditions (their simulations 1 through 4).  For each of these family forming simulations, we generate 1500 synthetic, evolved families using the procedure above.

\citet{Schlichting2009} did not perform collision simulations for the satellite breakup model, so we have to make some assumptions about the size and velocity distribution of the resulting family.  We adopt a differential size distribution for the family
\begin{equation}\label{eq:sd}
N(R)dR \propto R^{1-\beta}dR
\end{equation}
where $R$ is the radius of the collision fragments; we adopt values of $\beta$ in the range 4.5-5.5, consistent with typical catastrophic collision simulations \citep{Leinhardt2012,Marcus2011}.  Assuming a total family mass of $0.04-0.05 M_H$, we generate a synthetic family from this size distribution, with fragments in the size range $50$ km $ < R < 150$ km (the same size range as the graze-and-merge simulations, for ease of comparison).  We use the same density for the family members as for the graze-and-merge formation scenario to convert radius to mass.  For the ejection velocity of the fragments, we adopt a normal distribution with mean $200$~ms$^{-1}$ and standard deviation $50$~ms$^{-1}$.   We generate 1500 synthetic families using these assumptions and dynamically evolve the family members for 3.5 Gyr in the same way as described for the graze-and-merge model.  

\begin{figure} 
   \centering
   \includegraphics{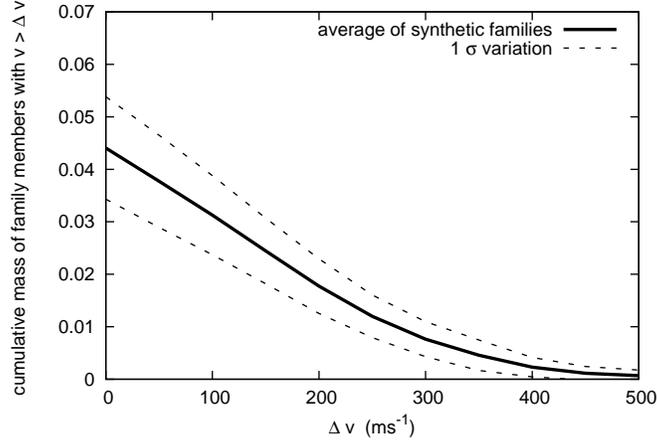}
   \caption{Average cumulative mass (in units of Haumea's mass) as a function of apparent $\Delta v$ at $t=3.5$Gyr (solid line) with $1\sigma$ variations (dashed lines) for 6000 synthetic graze-and-merge families (see section~\ref{ss:dynamical_evolution}).  For comparison, the total mass of the known Haumea family is estimated to be $\sim 0.017$ $M_H$ and is located mostly with $\Delta v < 150$~ms$^{-1}$.}
   \label{f:gm_mdv}
\end{figure}

\begin{figure} 
   \centering
   \includegraphics{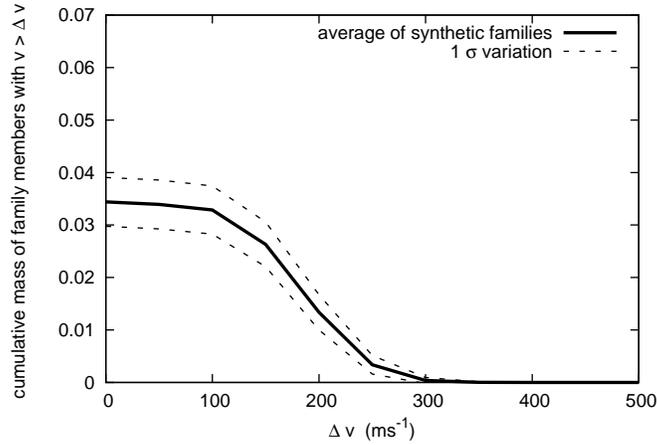}
   \caption{Average cumulative mass (in units of Haumea's mass) as a function of apparent $\Delta v$ at $t=3.5$Gyr (solid line) with $1\sigma$ variations (dashed lines) for 1500 synthetic satellite-breakup families families (see section~\ref{ss:dynamical_evolution}).}
   \label{f:sat_mdv}
\end{figure}

Figures~\ref{f:gm_mdv} and~\ref{f:sat_mdv} show our results for the \citet{Leinhardt2010} and \citet{Schlichting2009} formation models, respectively: we plot the average cumulative mass of the synthetic families as a function of their apparent $\Delta v$ at $t=3.5$ Gyr;   the $1\sigma$ uncertainties are also indicated.  Our calculations find that in the graze-and-merge model, the current evolved Haumea family should have a total mass of $0.045\pm0.01$ $M_H$, of which $\sim0.02$ $M_H$ should be found at $\Delta v < 150$~ms$^{-1}$.  The currently observed family (including Haumea's satellites) is estimated to have a mass of $\sim0.017$ $M_H$ \citep{Cook2011}, which accounts for $\sim85\%$ of the mass that is expected to be found within $150$~ms$^{-1}$ of the collision center for the \citet{Leinhardt2010} formation model.  The model predicts that there should be twice as much mass ($0.035\pm0.01$ $M_H$) in family members at larger velocities.  The satellite breakup model predicts a surviving family of $\sim0.035$ $M_H$, mostly in the $\Delta v = 100$--$300$~ms$^{-1}$ range, indicating that the known family members account for $\sim50\%$ of the expected mass of the family.

\subsection{Comparison with observations}\label{ss:observations}

The known Haumea family members are not an observationally complete population; to compare the evolved synthetic families from Section~\ref{ss:dynamical_evolution} to the observed family, we must estimate how many of our synthetic family members would be within the observable apparent magnitude and ecliptic latitude range of observational surveys that have detected the Haumea family members.  For each of our synthetic families (obtained in Section~\ref{ss:dynamical_evolution}), we take a snapshot of the instantaneous orbital elements of the family members at $t=3.5$ Gyr from their assigned orbital histories.  This snapshot allows us to calculate the heliocentric distance and ecliptic latitude for each synthetic family member.   To calculate the apparent magnitude, we use the heliocentric distance and the assigned size (as described in Section~\ref{ss:dynamical_evolution}), but we also need to make some assumptions about the albedos of the family members.  Haumea's albedo is 0.8 \citep{Lacerda2007}, and \citet{Elliot2010} have measured an albedo of 0.88 for the next brightest family member (2003 TX$_{300}$), but albedos have not been measured for the other family members.  Based on the light curves obtained for five of the known family members and the light curves of other icy solar system bodies with known albedos, \citet{Rabinowitz2008} argue that the Haumea family members' albedos are likely in the range $0.3-1.4$.  For our synthetic family, we adopt an albedo distribution with an average of 0.8 (Haumea's albedo) and a uniform spread of $\pm0.2$.  (We discuss below how this assumption might affect the results of our comparison.) Given this albedo assumption, we calculate the apparent magnitudes and ecliptic latitudes for each synthetic family, and we use the resulting distributions to calculate the number of objects  as a function of $\Delta v$ for each synthetic family that would be detected by an observational survey.

For the observational comparison, we use the results of the Palomar distant solar system survey conducted by \citet{Schwamb2010}.  This was a wide-field survey of $\sim12000$ $deg^{2}$ down to a limiting magnitude of $m_r\simeq21.3$, detecting 52 KBOs and Centaurs (27 previously known objects, and 25 new ones), including 4 of the previously identified Haumea family members.  The presence of so many known KBOs in their survey fields allowed \citet{Schwamb2010} to estimate that their detection efficiency was $\sim65\%$ down to $m_r\simeq21.3$ for the known population (see their Figure 3).  They also provide a plot of the survey's fractional sky coverage as a function of ecliptic latitude (see their Figure 4); they covered approximately $50\%$ of the sky $\pm30^{\circ}$ from the ecliptic.  From this information, we can estimate the detection probability for an object in our synthetic families based on its apparent magnitude and ecliptic latitude. We use this detection probability to determine how many of our synthetic collisional family members could have been detected in this survey.  
 
 An important note here is that the \citet{Schwamb2010} survey was not capable of spectrally identifying family members.  They can only say that there are four previously identified Haumea family members within their detections (listed in their Table 2), but it is possible that additional, unidentified Haumea family members are present within their survey detections.  To examine this possibility, we calculated $\Delta v$ for each of their listed detections and found two additional objects within $500$~ms$^{-1}$ of the collision center-of-mass orbit.  One of these objects, 2004 SB$_{60}$ ($\Delta v \approx 350$~ms$^{-1}$), was observed by \citet{Schaller2008} and found not to have the water ice spectral feature characteristic of all the other identified Haumea family members (its surface spectrum is consistent with no water ice being present).  The other object, 2008 AP$_{129}$, has a $\Delta v \approx 140$~ms$^{-1}$, suggestive of a dynamical association with the Haumea family, but the object shows only a moderate amount of water ice absorption in its surface spectrum \citep{Brown2012}, with a water ice fraction substantially lower than measured for the known family members.  If we allow for the possibility that 2008 AP$_{129}$ is a water ice poor family member, this means that it is possible that the \citet{Schwamb2010} survey detected as many as five Haumea family members, but this number is not likely to be larger.
 
\begin{figure} 
   \centering
   \includegraphics{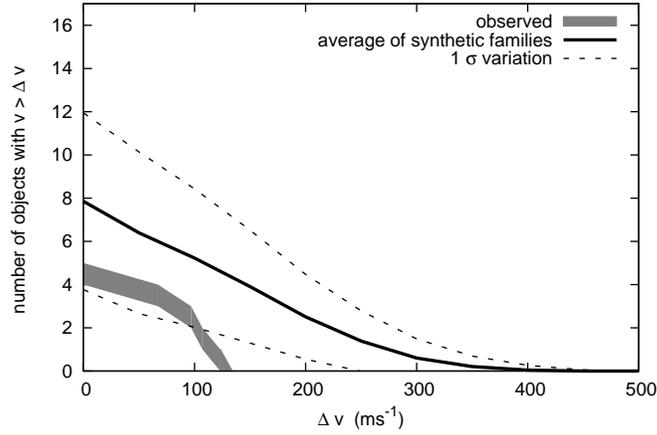}
   \caption{Cumulative velocity distribution of synthetic graze-and-merge family members observable by the \citet{Schwamb2010} distant solar system survey compared to the distribution of Haumea family members actually detected (see section~\ref{ss:observations}).}
   \label{f:ndv_gm}
\end{figure}

\begin{figure} 
   \centering
   \includegraphics{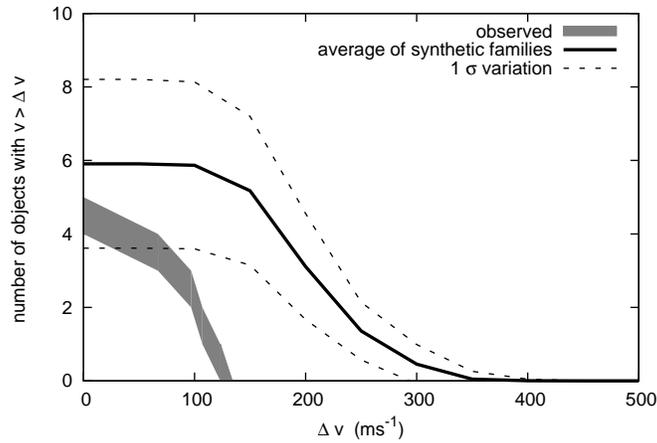}
   \caption{The same as for Figure~\ref{f:ndv_gm} but for the satellite-breakup formation model.}
   \label{f:ndv_sat}
\end{figure}

Figures~\ref{f:ndv_gm} and~\ref{f:ndv_sat} show the number of synthetic family members as a function of $\Delta v$ for each of the two formation scenarios that would have been detected by the \citet{Schwamb2010} survey; also shown for reference are the Haumea family members actually detected in that survey.

For the satellite breakup model (Figure~\ref{f:ndv_sat}), the actual number of family members detected falls within the $1 \sigma$ range of total detections predicted by the model, but the observed values of $\Delta v$ are lower than the predicted values.  Almost none of the synthetic families match the observations by producing 4 or 5 detections all with $\Delta v < 150$~ms$^{-1}$.  The values of $\Delta v$ for some of the observed family members could be larger than the values in Table~\ref{t:known_family} though, because of the uncertainty in the orbital angles of the collision orbit.  To constrain the collision orbit and calculate the minimum $\Delta v$ for the known family members, we assumed that the collision took place near the ecliptic plane (where collision probabilities are highest);  the family members' ejection velocities from the collision could be larger than this minimum value, although \citet{Ragozzine2007} argue that the correction factor is likely to be $\sim2$ or lower unless the ejection of fragments was highly anisotropic (see also our discussion of this in Section~\ref{ss:cmorbit}).  If we allow for a factor of two correction to the $\Delta v$ estimates for the real family members, we can extend the $\Delta v$ of the survey's detected family members out to $\sim250$~ms$^{-1}$.  Using this increased allowable range of $\Delta v$, $18\%$ of the synthetic families satisfactorily reproduce the observations, making the satellite breakup model statistically consistent with the observations.

The graze-and-merge model predicts that, on average, the survey should have detected 8 family members.  This is larger than the 4 or 5 real detections, but the actual detections fall within the $1\sigma$ uncertainties of the synthetic families.  Just as in the satellite breakup model, the real detections fall at significantly lower $\Delta v$ than predicted by the graze-and-merge model.  Very few of the synthetic families result in all the detectable family members falling below $150$~ms$^{-1}$, and all of these cases result in too few detections to be consistent with the observations.  If we increase the allowed maximum $\Delta v$ to $250$~ms$^{-1}$ and require the synthetic families to match the number of real detections (4 or 5), we find that only $4\%$ of the synthetic families reproduce the observations, indicating that the observations are not a typical outcome of the graze-and-merge model.  We did, however, make a number of assumptions when generating our synthetic families, so we examine how these might be altered to bring the model into closer agreement with the observations.

One assumption we made in the creation of the families in Section~\ref{ss:dynamical_evolution} was that there was no relationship between a fragment's mass and its $\Delta v$ from the collision center.  Our only constraint was that the binned mass vs. $\Delta v$ matched the outcome of the \citet{Leinhardt2010} simulations.  If there is a correlation between fragment mass and $\Delta v$ such that the higher $\Delta v$ family members are, on average, smaller than the lower $\Delta v$ members, this might account for the lack of observed family members at large $\Delta v$.  To test this we impose a relationship between fragment mass, $m$, and $\Delta v$ such that, averaged over all the family members,
\begin{equation}\label{eq:mdv}
\Delta v(m) \propto m^{-k}
\end{equation}
while the total mass of all fragments within a given range of $\Delta v$ is still constrained by the \citet{Leinhardt2010} simulation results.  This power law relationship has been seen in some laboratory impact experiments, with the exponent typically being $k < 1/6$ \citep{Nakamura1991,Holsapple2002,Giblin2004}.  Taking $k=1/6$ and generating a new set of synthetic families for the graze-and-merge collision, the percentage of synthetic families with 4 or 5 detections all with $\Delta v < 250$~ms$^{-1}$ increases to $8\%$.  The percentage of synthetic families with 4 or 5 detections and $\Delta v < 150$~ms$^{-1}$ is still negligible; even if we increase the value of $k$ to $1/4$ (which is not a likely value), we still see $<1\%$ of the synthetic families resulting in 4 or 5 detections below $150$~ms$^{-1}$.

Another easily altered assumption is that of the albedos for the family members.  We assumed albedos in the range $0.6-1.0$, but if we assume systematically lower albedos, in the range $0.4-0.6$ (still consistent with their icy composition), we decrease the average predicted number of detections for the graze-and-merge model from $\sim8$ to $\sim4$, also decreasing the number of predicted detections at large $\Delta v$.  With this albedo assumption, nearly half of the synthetic families result in no detections at $\Delta v > 250$~ms$^{-1}$, and the percentage of families with 4 or 5 detections and $\Delta v < 250$~ms$^{-1}$ is $11\%$.  Lowering the albedos further does not increase the percentage of families that agree with the observations.  If we assume the above albedo range in combination with the $k=1/6$ mass-$\Delta v$ relationship, the percentage of matching families is $\sim12\%$.  Given that at least one family member besides Haumea has been shown to have a very bright albedo \citep{Elliot2010}, assuming systematically lower albedos for most of the family members might not be the most likely solution to the discrepancy between the observations and the models.  However, it cannot be ruled out until the albedos of some of the smaller family members have been measured.

Depending on the assumptions about albedos and mass-velocity relationships, and allowing for the uncertainty in the known family's $\Delta v$, we find that $4-12\%$ of the synthetic families generated by the graze-and-merge collision model reproduce the Haumea family, as observed by the \citet{Schwamb2010} survey.  The actual observed family is at systematically lower $\Delta v$ than the model predicts, but given the relatively small set of observed family members, the model is still consistent the observations.

\section{Discussion and Conclusions}\label{s:conclusions}

After accounting for 3.5 Gyr of dynamical evolution and accounting for the observational incompleteness of the known family, both of the proposed Haumea family formation models we have examined here \citep{Leinhardt2010,Schlichting2009} are consistent with the total number of observed family members.  There is, however, a significant difference between the observed values of $\Delta v$ and those predicted in the formation models; almost none ($\ll1\%$) of the synthetic families we generated for either formation scenario account for the observations of family members that all fall within $150$~ms$^{-1}$ of the collision center.  The only way we find to make the velocity distributions for the formation scenarios consistent with the observations is to allow that the actual ejection velocities of some of the known family members are a factor of $\sim2$ larger than the calculated minimum values.  This adjustment of the observed $\Delta v$ falls within the uncertainties for their calculation of the collision center-of-mass orbit (see our discussion in Section~\ref{ss:cmorbit}). But even with this adjustment, only $10-20\%$ of the synthetic families reproduce the observed family.  This is a reasonable level of agreement given the uncertainties in the collision models and the small number of observed family members, but it is still interesting that so few family members have been identified at large $\Delta v$.  In section~\ref{ss:observations} we explored how the assumptions we made about the albedos of the family members and possible correlations between fragment mass and ejection velocity could change the predictions of the graze-and-merge collision model.  We find that it is difficult to alter these assumptions enough to obtain a better than 10--15\% agreement with the observations.  For either formation scenario, we find that there should be an additional $\sim0.01-0.03$ $M_H$ of family members that have yet to be identified, i.e., at least as many as those already detected.  If, as we discover these additional family members, the distribution of $\Delta v$ remains too heavily weighted toward the low-end, the formation models will have to be reconsidered.

\begin{figure} 
   \centering
   \includegraphics{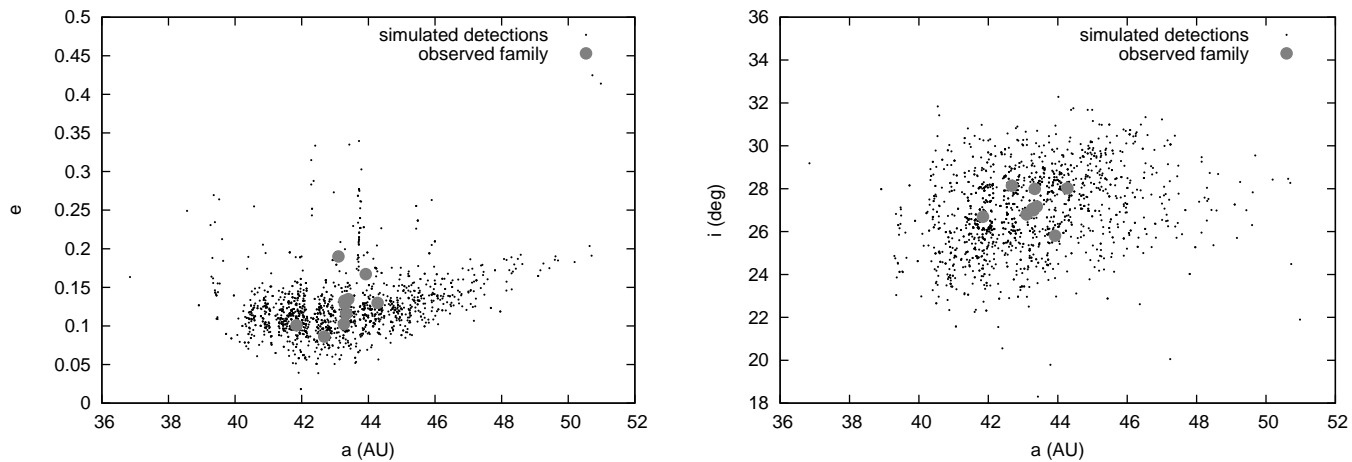}
   \caption{Eccentricity vs. semimajor axis (left panel) and inclination vs. semimajor axis (right panel) for the simulated detections (black dots) from a representative subset of the synthetic graze-and-merge collisional families.  The observed family is shown in grey.}
   \label{f:simulated_detections}
\end{figure}

Recent spectroscopic and photometric surveys of the Kuiper belt designed to detect water ice have not identified any large $\Delta v$, ice-rich Haumea family members.  \citet{Fraser2012} performed a photometric study of 120 objects from the major dynamical classes of the Kuiper belt using the Hubble Space Telescope (HST) and did not find any additional Haumea family members.  \citet{Benecchi2011} also performed HST photometry of a large sample of Kuiper belt objects and failed to identify any new ice-rich family members.  Ground based studies searching for water ice in the Kuiper belt have also failed to detect higher $\Delta v$ family members; \citet{Brown2012} detected no additional family members, and \citet{Trujillo2011} found only one new member, located near the dynamical core of the family.  The extent of these surveys suggests that if there were ice-rich Haumea family members spread at large $\Delta v$ throughout the Kuiper belt, we likely should have already identified some of them.  This is consistent with our comparisons  of the proposed collisional models to the observed Haumea family; figure~\ref{f:simulated_detections} shows the distribution of simulated detections for a subset of our dynamically evolved graze-and-merge collisional families (see Section~\ref{ss:observations}) compared to the known Haumea family.  The simulated detections span a larger parameter range of the Kuiper belt (due to the presence of higher $\Delta v$ family members) than the actual detections;  \citet{Lykawka2012} also found that simulated Haumea families with $\Delta v$ consistent with existing collisional formation models would occupy a larger portion of the Kuiper belt than currently observed.   In contrast to the expected $\Delta v$ distributions from the collisional models, we find that, accounting for the time evolution of the a-e-i and $\Delta v$ distributions in our simulations, all of the known family members are consistent with initial $\Delta v < 100$ ms$^{-1}$.

One possible way to modify the collision models was suggested by \citet{Cook2011}, who argue that perhaps the proto-Haumea was only partially differentiated.  This would allow for some of the collisional fragments to be rocky rather than primarily water ice, which would mean that they might not show the water ice spectral feature that has thus far been used as the only secure way to identify a family member; perhaps the dynamically nearby KBO 2008 AP$_{129}$, which has a $\Delta v$ of only 140~ms$^{-1}$ but shows less water ice absorption than the accepted family members (see Section~\ref{ss:observations}) represents a new class of rockier family members.  A rockier composition could also make the fragments darker and therefore more difficult to detect at all.  It has been similarly suggested that surface inhomogeneities on the proto-Haumea could result in collisional fragments with different compositional characteristics from those of the known family members \citep{Schaller2008}.  It is unclear why, in either of these situations, there would be a preference for the less icy fragments to be dispersed at large $\Delta v$, but if the composition of the target and impactor are substantially different from those assumed in the collision simulations, that could affect the entire $\Delta v$ distribution.  Another possibility is that the formation models have not adequately accounted for collisional evolution amongst the ejected fragments themselves, and that this could alter the family's size and/or velocity distribution in a significant way.  Collision simulations like the ones in \citet{Leinhardt2010} are computationally very expensive and they follow the family's evolution for only a few thousand spin periods of the primary (a few hundred days in total), so the model's final size and velocity distributions are not fully evolved.  

The presence or absence of higher $\Delta v$ family members with future additions to the set of observed Haumea family members will determine if any of these modified scenarios should be considered, or if there is another collisional model that could better explain the family.  For either the \citet{Leinhardt2010} or the \citet{Schlichting2009} models to be consistent with observations, we should find several higher $\Delta v$ family members among any new identifications.

In summary, our study of the long-term dynamical evolution of the Haumea family leads to the followings conclusions.

\begin{enumerate}

\item The Haumea family is at least 100 Myr old.  This estimate is based on the timescale to randomize the nodal longitudes of the orbital planes of the family members, as well the timescale for chaotic evolution of Haumea's eccentricity in the 12:7 MMR with Neptune.  From the chaotic diffusion of Haumea's eccentricity, we can conclude with $95\%$ confidence that the family is older than 1 Gyr.

\item For initial ejection velocities, $\Delta v$, in the range $50-400$~ms$^{-1}$, $20-45$\% of original Haumea family members are lost due to close encounters with Neptune over 3.5 Gyr.  Most of this loss occurs at the inner edge of the family (interior to $\sim 41$ AU) and near the locations of MMRs with Neptune.  A few percent of the surviving Haumea family members are expected to be found in MMRs with Neptune.  The 3:2 and 7:4 MMRs are the most likely of the resonances to contain surviving members.

\item Within the population of surviving and potentially recognizable family members, chaotic diffusion in orbital elements over 3.5 Gyr introduces a $50-100$~ms$^{-1}$ spread in the apparent velocities of the family relative to the collision center-of-mass orbit, with the average $\Delta v$ increasing slightly over time.

\item Accounting for long-term dynamical evolution to the graze-and-merge collision model of \citet{Leinhardt2010}, we find that the currently observed family represents $>85\%$ of the expected family mass within $150$~ms$^{-1}$ of the collision center, but an additional $0.035\pm0.01$ $M_H$ (about twice the mass of the known family) remains to be identified at larger $\Delta v$.  Accounting for observational incompleteness, the \citet{Leinhardt2010} model is consistent with the observations at the $\sim10\%$ confidence level.

\item For the satellite breakup model of \citet{Schlichting2009}, we find that the currently observed family accounts for $\sim 50\%$ of the expected mass of the family.  Most of the remaining mass should be found at $\Delta v > 150$~ms$^{-1}$.  Accounting for observational incompleteness, the satellite breakup model is consistent with the observations at the $\sim20\%$ confidence level.

\item Both formation models predict more family members at large $\Delta v$ than are currently observed (even allowing for a factor of $\sim2$ higher values of $\Delta v$ for the known family members due to the uncertainty in estimates of the collision center-of-mass orbit).  If additional Haumea family members are identified and continue to have low $\Delta v$ ($\lesssim 200$~ms$^{-1}$), new formation models (or modifications to the existing models) will have to be considered. 

\end{enumerate}

\acknowledgments
This research was supported by grant no.~NNX08AQ65G from NASA's Outer Planets Research program.  
We thank D.~Ragozzine for a helpful review.


\clearpage

\begin{deluxetable}{l l l l l l l}
\tabletypesize{\footnotesize}
\tablecolumns{7}
\tablewidth{0pt}
\tablecaption{Known Haumea Family Members}
\tablehead{ \colhead{Designation} & \colhead{a (AU)} & \colhead{e} & \colhead{i (deg)} & \colhead{$\Delta v_1$ (ms$^{-1}$)} & \colhead{$\Delta v_2$ (ms$^{-1}$)$^{B}$} & \colhead{absolute magnitude} }

\startdata
Haumea (2003 EL$_{61})$	& 43.10	& 0.19	& 26.8	& 320	& 342	& 0.27	 \\ 
1995 SM$_{55}$			& 41.85	& 0.10 	& 26.7 	& 158	& 174	& 6.3	\\
1996 TO$_{66}$			& 43.34	& 0.12	& 28.0	& 25		& 46		& 4.5	\\
1999 OY$_3$				& 43.92	& 0.17	& 25.8 	& 293	& 272	& 6.76	\\
2002 TX$_{300}$			& 43.30	& 0.13  	& 27.0	& 107	& 97		& 3.09	\\
2003 OP$_{32}$			& 43.25 	& 0.10	& 27.0 	& 120	& 138	& 4.2	\\
2003 UZ$_{117}$			& 44.28	& 0.13	& 28.0	& 67		& 60		& 5.2	\\	
2005 CB$_{79}$			& 43.40 	& 0.13  	& 27.2 	& 111	& 94		& 5.4	 \\
2005 RR$_{43}$			& 43.28	& 0.13	& 27.1 	& 113	& 98		& 4.0	\\
2003 SQ$_{317}$			& 42.68	& 0.09	& 28.1 	& 144	& 172	& 6.3	\\
\enddata

\tablecomments{$\Delta v_1$ is calculated using the \citet{Ragozzine2007} center-of-mass orbit and $\Delta v_2$ is calculated using our updated center-of-mass orbit (see section~\ref{ss:cmorbit}).  When calculating $\Delta v$, $a$, $e$, and $i$ for the family members are fixed at their current values, but the orbital angles are allowed to vary freely.}

\label{t:known_family}
\end{deluxetable}

\begin{deluxetable}{l l l l l l}
\tabletypesize{\footnotesize}
\tablecolumns{3}
\tablewidth{0pt}
\tablecaption{Fraction of simulated test particles that survive to 1.5 and 3.5 Gyr}
\tablehead{\colhead{$\Delta v$ (ms$^{-1}$)} & \colhead{t = 1.5 Gyr} & \colhead{t = 3.5 Gyr}}

\startdata
50	& 0.89	& 0.81\\
100	& 0.87	& 0.77\\
150	& 0.88	& 0.80\\
200	& 0.79	& 0.74\\
250	& 0.75	& 0.67\\
300	& 0.70	& 0.64\\
350	& 0.65	& 0.61\\
400	& 0.63	& 0.57\\
\enddata


\label{t:survival_rates}
\end{deluxetable}

\end{document}